\documentclass[twocolumn,showpacs,preprintnumbers,amsmath,amssymb,prb]{revtex4}

\usepackage{graphicx}
\usepackage{dcolumn}
\usepackage{bm}

\newcommand{\be}{\begin{equation}}
\newcommand{\ee}{\end{equation}}
\newcommand{\bea}{\begin{eqnarray}}
\newcommand{\bean}{\begin{eqnarray}\nonumber}
\newcommand{\eea}{\end{eqnarray}}
\newcommand{\Eq}[1]{Eq.\,(\ref{#1})}
\newcommand{\Sect}[1]{Sect.\,(\ref{#1})}
\newcommand{\App}[1]  {App.\,(\ref{#1})}
\newcommand{\Fig}[1]{Fig.\,\ref{#1}}
\newcommand{\Tab}[1]{Tab.\,\ref{#1}}
\newcommand{\bfi}{\begin{figure}}
\newcommand{\efi}{\end{figure}}
\newcommand{\bmi}{\begin{minipage}}
\newcommand{\emi}{\end{minipage}}

\newcommand{\half}{\frac{1}{2}}

\newcommand{\q}{{\mathbf q}}
\newcommand{\qp}{{\q_\parallel}}

\newcommand{\R}{{\mathbf R}}

\newcommand{\ro}{{\mbox{\boldmath $\rho$}}}
\newcommand{\roe}{{\mbox{\boldmath $\rho$}}_e}
\newcommand{\roh}{{\mbox{\boldmath $\rho$}}_h}

\newcommand{\DeltaR}{\Delta_{\mathbf R}}

\newcommand{\wq}{w_{\mathbf q}}

\newcommand{\ea}{\epsilon_{\alpha}}

\newcommand{\eb}{\epsilon_{\beta}}

\newcommand{\psia}{\psi_\alpha(\mathbf{R})}
\newcommand{\psib}{\psi_\beta(\mathbf{R})}

\newcommand{\Ba}{B^{\vphantom {\dagger }}_{\alpha}}
\newcommand{\Bad}{B^{\dagger}_{\alpha}}
\newcommand{\Bb}{B^{\vphantom {\dagger }}_{\beta}}
\newcommand{\Bbd}{B^{\dagger}_{\beta}}
\newcommand{\tabq}{t_{\alpha \beta}^{\q}}

\newcommand{\etal}{{\em et al.\,}}
\newcommand{\inte}{\int\!}

\renewcommand{\dag}{^{\dagger}}

\renewcommand{\r}{{\mathbf r}}

\begin{document}


\title{Near-field spectra of quantum well excitons \\
with non-Markovian phonon scattering}

\author{G. Mannarini}
 \altaffiliation[Also at ]{Institut f\"ur Physik der Humboldt-Universit\"at
   zu Berlin, Newtonstr. 15, 12489 Berlin, Germany}
\affiliation{National Nanotechnology Laboratory of CNR, via
Arnesano 16, 73100 Lecce, Italy}
 \email{Gianandrea.Mannarini@unile.it}
\author{R. Zimmermann}%
\affiliation{Institut f\"ur Physik der Humboldt-Universit\"at
   zu Berlin, Newtonstr. 15, 12489 Berlin, Germany}%

\date{\today}

\begin{abstract}
The excitonic absorption spectrum for a disordered quantum well
in presence of exciton-acoustic phonon interaction is treated
beyond the Markov approximation. Realistic disorder exciton
states are taken from a microscopic simulation, and the
deformation potential interaction is implemented. The exciton
Green's  function is solved with a self energy in second order
Born approximation. The calculated spectra differ from a superposition
of  Lorentzian lineshapes by enhanced inter-peak absorption. This
is a manifestation of pure dephasing which should be possible to
measure in near-field experiments.
\end{abstract}

\maketitle

\section{\label{sec:introduction} Introduction}

Unique information on semiconductor nanostructures  even beyond the
diffraction limit can be
obtained using near-field spectroscopy. This opens the way to detect fine
spectroscopic and structural details \cite{runge05}. Usual
selection rules for optical transitions are broken, which allows one
to observe also states which are dark in the far-field
\cite{hohenester05}. The present record experiment is a spatial
resolution of $30$\, nm which  was achieved using a tapered
optical fiber and an InAs self-assembled quantum dot (QD) sample
\cite{matsuda02}. Even with a less demanding setup using a resolution
 comparable to the wavelength of light, individual optical transitions
 can be seen due to their spatial and spectral position.
 Absorption spectra from quantum wells (QW)
with interface fluctuations could be measured in transmission
geometry through a metal aperture of $400$\, nm on a GaAs/AlGaAs
sample whose substrate was removed  \cite{guest02}. This gives
access to single exciton state absorption lineshapes, with the
chance to detect deviations from a Lorentzian profile. Similar
lineshapes are seen in photoluminescence (PL). Among others,
Besombes \etal  \cite{besombes01} found that the PL from a single
CdTe QD exhibits broad bands on the sides of the Lorentzian peak.
This has been also seen in Fourier-transformed four wave mixing
experiments by Borri \etal \cite{borri01,borri05}, who
systematically studied the importance of the broad band with
respect to the (Lorentzian broadened) zero phonon line (ZPL) as a
function of confinement energy in InAs/InGaAs QDs.

However, using the standard Fermi's golden rule for the phonon
scattering, only a Lorentzian broadened ZPL is expected for the
absorption spectrum of a single exciton state.  Phenomena which
go beyond the framework of  Fermi's golden rule (such as the phonon
broad band) are often called quantum kinetic or non-Markovian
(i.e. memory) effects. The broad band can be understood in terms
of a non-perturbative coupling between electronic and phonon
degrees of freedom. This goes back to the  Huang-Rhys theory of
$F$-centers \cite{duke65}, which considers intra-state (diagonal)
coupling to phonons only. This problem can be solved exactly with
the cumulant expansion (also  known as  Independent Boson Model
\cite{mahan}). Since off-diagonal level coupling is absent, the
ZPL remains unbroadened. Within this model, different
exciton-phonon couplings such as  optical polar, acoustic
piezoelectric, and deformation potential have been studied
\cite{bondarev03}. This model was recently extended to include
off-diagonal coupling, successfully explaining why the ZPL get
broadened  even if real phonon-assisted transitions to other
levels are not possible (pure dephasing)
\cite{muljarov04,muljarov05}. Apart from the cumulant expansion,
approaches based on various dynamical approximations have been
used, too. For instance, a 4th order selfconsistent summation of
diagrams allowed one to compare QW linewidths extracted from photon
echo measurements \cite{fan98, tak99}: It was found that at
elevated temperatures the linewidth broadening is mainly due to
pure dephasing.  However, a simple quantum disk model for
computing the exciton wave functions was used in this case. The
correlation expansion for the carrier-phonon interaction which
was used in Ref.\onlinecite{krum02} to model the coupling to both
optical and acoustical phonons, is exact up to the 2nd order.
However, neither Coulomb or disorder effects were considered
there.

A consistent theoretical treatment which describes non-Markovian
effects for {\em realistic}, disordered excitonic resonances from
semiconductor QWs is, to the best of our knowledge, still
missing. In this paper we face this problem accounting for both
diagonal and off-diagonal coupling to acoustic phonons via
deformation potential interaction. The excitonic self energy is
taken in second Born (2B) approximation and integrated numerically.
This approach reproduces both the real scattering processes
leading to the major ZPL broadening and the level-diagonal pure
dephasing responsible for the broad band. The resulting
(near-field) absorption  spectrum exhibits apart from the
Lorentzian-broadened individual lines a superposition of broad
bands which results in an enhanced inter-peak absorption. We
suggest that this effect of non-Markovian dynamics can be
demonstrated in near-field measurements.

The paper is organized as follows: The theory with a presentation
of different approximations for the self energy in
\Sect{sec:theory} is followed by  numerical results in
\Sect{sec:results}, and the conclusions are given in
\Sect{sec:conclusions}. Finally, two appendixes contain more
technical details.

\section{\label{sec:theory} Theory}

To model a realistic QW with disorder it is convenient
to start with disorder eigenstates of the exciton. If the
disorder strength is much smaller than both the level distance of
electron and hole confinement states and the exciton binding
energy, the total exciton wave function (in envelope function and
effective mass approximation) can be factorized as \cite{zim97}
\be          \label{FactorWF}
\Psi_\alpha (\r_e,\r_h) =
  u_e(z_e) \, u_h(z_h) \, \phi_{1s}(\roe-\roh) \, \psia \, ,
\ee
where confinement $u_a(z_a)$ in growth direction $z$,  in-plane
exciton relative motion $\phi_{1s}(\ro)$, and  in-plane
center-of-mass (COM) wavefunction $ \psia$ appear. Since the
first two functions are independent of disorder, they have to be
computed just once for the disorder-free QW. Then, it
suffices to solve a single particle equation for the COM exciton
states in two dimensions,
\be          \label{COMmotion}
\left[ - \frac{\hbar^2}{2M} \DeltaR + V(\R) \right] \psia =
\ea \, \psia \, .
\ee
The potential $V(\R)$ accounts for interface disorder and is
spatially correlated on the scale of the exciton relative motion,
\be\label{V_def}
V(\R) =
  \int \! d\R' \sum_{a=e,h} \eta_a^2 \, \phi_{1s}^2[\eta_a(\R-\R')] \,
 \frac{dE_a}{dL_z} \, \Delta L_z(\R') \, ,
\ee
with mass ratios $\eta_e = M/m_h$ and $\eta_h = M/m_e$. The well
width fluctuations $\Delta L_z(\R)$ transfer into potential
fluctuations via the derivative of the ideal electron and hole
confinement energy $dE_a/dL_z$ \cite{zim_rev03}.
As zero of energy, we take the ideal 1s exciton transition energy
of the lowest sublevel transition hh1-e1 in a
GaAs/Al$_{0.3}$Ga$_{0.7}$As  QW of given width $L_z$.

Recently, Grochol  \etal \cite{grochol05} have critically
questioned the factorization Ansatz Eq.(1). Calculations with the
full in-plane electron-hole wave function gave systematically
lower eigenergies, while the wave functions  showed only minor
modifications compared to the factorization, at least in the
low-energy part of the spectrum. Since in the present paper, we
are interested in the low-energy tail of the absorption spectrum
with well localized states, the factorization can be safely used.

The interaction with acoustic phonons is treated here as the main
source of scattering and dephasing. Optical phonon scattering
is not important here since, at temperatures below room
temperature, optical phonon absorption by the exciton states is
unlikely. Further, piezoelectric scattering is neglected since
the exciton wave functions are typically large in momentum space,
leading to small matrix elements for this kind of coupling
\cite{run_rev02, tak93}. Thus, only deformation potential
interaction with longitudinal acoustic phonons is relevant.

This model leads to the following Hamilton operator for excitons
(creation operators $\Bad$) and acoustic phonons (creation
operators $a_\q\dag$ with three-dimensional momentum $\q$) with
(bulk like) dispersion $\wq=u_S\,|\q|$) :
\bean     \label{ham} {\cal H} & = & \sum_\alpha \ea \, \Bad \Ba
\,        + \, \sum_{\q} \hbar\wq \, a_\q\dag a_\q   \\
& + & \sum_{\alpha \beta \q} \tabq (a_\q\dag + a_{-\q} ) \Bad \Bb    \,.
\eea
The deformation potential matrix elements $\tabq$ are discussed
in \App{sec:app_me}. We have neglected the spin degree of freedom
which splits each localized exciton state into a bright doublet
via the exchange interaction. This splitting is driven by COM
anisotropy and has typical values of a few tens of $\mu$eV
\cite{zim_rev03}. It can be disregarded in calculating absorption
spectra with not too high resolution.

Within linear response theory, the optical properties of the
system are given by the one-exciton Green's function (exciton
propagator)
\be G_{\alpha\beta}(t -t')= -i\langle{\cal T}\Ba (t) \Bbd(t') \rangle \ee
with time ordered finite-temperature expectation value. In
frequency space, the Green's function obeys the Dyson equation
(in what follows $\hbar=1$ is used, and the frequency $\omega$ is
understood to carry a small negative imaginary part, $\omega - i
0$)
\be\label{lin_sys} \sum_{\eta} \left[ ( \omega -
\ea)\delta_{\alpha\eta}  - \Sigma_{\alpha\eta}(\omega) \right]
G_{\eta\beta} (\omega) = \delta_{\alpha\beta} \, , \ee
introducing the exciton self energy matrix
$\Sigma_{\alpha\eta}(\omega)$. The absorption spectrum is
proportional to
\be \alpha(\omega) = \mathrm{Im}\sum_{\alpha \beta}m_\alpha^*\,
G_{\alpha\beta}(\omega)\, m_\beta \, . \ee
The optical interband matrix elements $m_\alpha$ are defined in
\App{sec:app_me}.

For the self energy, we take the lowest order diagram for the
exciton-phonon interaction shown in \Fig{im:fig1}. It is called
2nd order Born (2B) approximation since the interaction appears
just twice:
\be \label{self_basic} \Sigma_{\alpha\eta}(\omega) =
\sum_{\beta\q} \tabq t_{\beta \eta}^{-\q} \left[ \frac{1 +
n(\wq)}{\omega - \eb - \wq} +  \frac{ n(\wq)}{\omega - \eb + \wq}
\right] \, . \ee
The phonons are assumed to be in equilibrium at the lattice
temperature $T$, $n(\wq)=1/(\mathrm{exp}(\wq/k_B T)-1)$. The two
terms in \Eq{self_basic}  are related to phonon emission and
absorption. Note that the internal line in \Fig{im:fig1} is
understood to represent the bare exciton propagator. Taking
instead a dressed propagator with shifted quasiparticle energy
would be an improvement. Our calculations have shown that this
(acoustic) polaron shift for the localized exciton states is of
the order of a few $\mu$eV only, and we have chosen to neglect it
everywhere.

An alternative derivation would start with the equations of
motion for the exciton propagator which couples to mixed
exciton-phonon operator expectation values (phonon-assisted
density matrix). The next equation in the hierarchy is then
decoupled in a way that two phonon operators can be combined into
the phonon occupation function. Eliminating the phonon assisted
density matrix from the equations leads to a differential
equation of the exciton propagator with a scattering being
essentially a time integration over the past. This kind of memory
in the scattering is fully equivalent to the frequency dependence
of the self energy in the diagrammatic approach. Further, the
non-diagonality of the self energy  $\Sigma_{\alpha\eta}(\omega)$
leads to cross correlations among individual excitons.

In the Markov approximation these correlations are neglected, but
more importantly the frequency dependence of the self energy is
dropped completely. In the time frame, this means to extract the
exciton propagator from the scattering integral at the latest
time, thus skipping any memory effect. Consequently, the temporal
dynamics of the propagator reduces to an exponential decay with a
constant damping $\gamma_\alpha^M$. In this paper we go beyond
this level of approximation, aiming at a full non-Markovian
description which includes the non-diagonality and the memory
(frequency dependence) in the self energy. For the sake of
comparison, we discuss in the following  subsections the solution
of \Eq{lin_sys} at various levels of sophistication.

The radiative contribution to the self energy (radiative damping)
will be neglected since it is tiny in thin QWs which have
strongly localized excitons \cite{zim_rev03}.

\subsection{Full non-Markovian level}\label{sec:theory_nm}

The self energy \Eq{self_basic} can be written in a more compact
form by extending the energy integration to positive and negative
values,
\be \label{self_def} \Sigma_{\alpha\eta}(\omega) = \sum_\beta\int\!dE\,
n(E)\, \frac{J_{\alpha\eta}^\beta(E)}{\omega - \eb + E}  \, ,
\ee
which accounts for both phonon absorption and emission. The
coupling function
\be \label{J_def} J_{\alpha\eta}^\beta(E) \equiv \mathrm{sgn}(E)
\sum_\q  t_{\alpha\beta}^\q t_{\beta\eta}^{-\q}   \,  \delta(|E|
- \omega_\q) \,. \ee
can be viewed upon as a phonon density of states weighted by
coupling matrix elements. Its spectral range
decides which exciton states are coupled to the phonon bath. It
starts at least with a power of $E^3$ at small energies, as seen
in \App{sec:app_me}. Since the exciton wave functions are real valued, the
following symmetries hold: $J_{\alpha\eta}^\beta(E) =
J_{\eta\alpha}^\beta(E)$ and $J_{\alpha\alpha}^\beta(E) =
J_{\beta\beta}^\alpha(E)$. Thus, if we consider a system of $N$ exciton
states, $(N^3+N)/2$ components of the coupling function have to
be computed.
\bfi[t]
\centering
\includegraphics*[width=5cm]{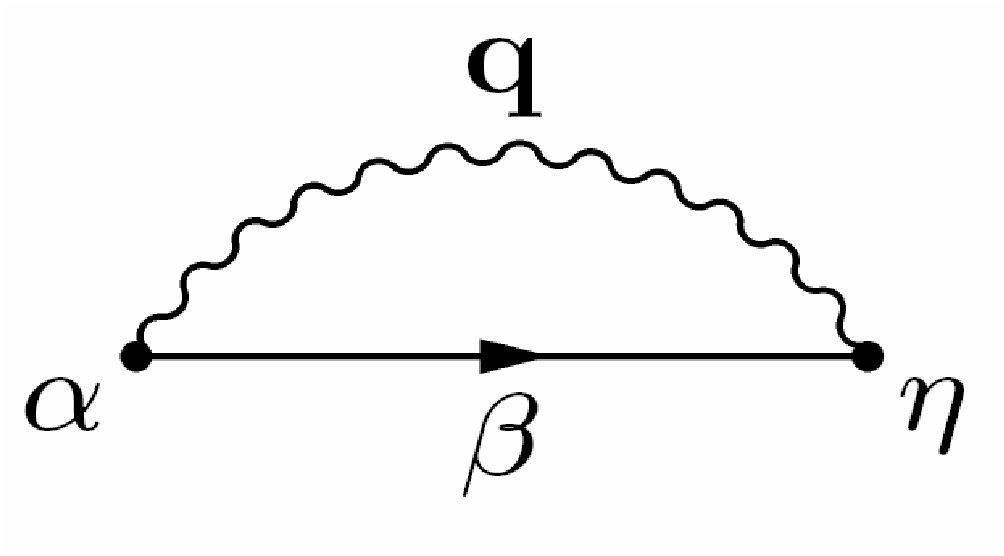}
\caption{\label{im:fig1}Second order self energy diagram. The
arrow denotes the exciton propagator, while the wiggly line
stands for the phonon propagator. Each dot carries a coupling
element $t$.} \efi

\subsection{Markov approximation}

The Markov approximation for the exciton-phonon dynamics is
defined within our formalism by keeping only diagonal terms of
the self energy and evaluating them at the exciton resonance
(on-shell),
\be \label{M_self} \Sigma_{\alpha\eta}(\omega) \, \rightarrow \,
\delta_{\alpha\eta} \Sigma_\alpha^{M}\, , \quad \Sigma_\alpha^{M}
\equiv \Sigma_{\alpha\alpha} (\omega=\ea-i0) \,. \ee
Apart from the (small) polaron shift originating from the real
part, this level of approximation implies a constant damping
\be \label{markov_gamma} \gamma^M_\alpha/2 = \mathrm{Im}
\Sigma_\alpha^{M} = \pi \sum_{\beta} n(\eb-\ea) \,
J_{\alpha\alpha}^\beta(\eb-\ea) \, . \ee
(The factor 2 appears because $\gamma^M_\alpha$ is the decay rate
of the intensity, which is proportional to the squared exciton
propagator.) Thus, \Eq{lin_sys} reduces to a simple algebraic equation,
and the absorption comes out as
\bean
\alpha(\omega) &=& \sum_\alpha |m_\alpha|^2 \mathrm{Im}
\frac{1}{\omega  - \ea - i\gamma^M_\alpha/2} \\
& = & \sum_\alpha |m_\alpha|^2  \frac{\gamma^M_\alpha/2}{(\omega
- \ea)^2 + (\gamma^M_\alpha/2)^2} \, , \eea
which is a Lorentzian lineshape for each line. No broad bands are
obtained within the Markov approximation. The rate
$\gamma^M_\alpha$ is identical to the result of Fermi's golden
rule. We notice that within the Markov approximation, only
$(N^2+N)/2$ components of the coupling function have to be
computed.

\subsection{Single state limit}

As already mentioned, it is not possible to account for broad
bands within the Markov approximation. In the single-state limit
(uncoupled exciton states), the Markov rates \Eq{markov_gamma}
vanish since the coupling function $J_{\alpha\alpha}^\alpha(E)$
tends strongly to zero at zero argument $E = \ea - \ea$.  Thus,
in the absence of radiative damping, the absorption consists of
delta lines. However, if we do not perform the Markov
approximation, a composite absorption lineshape is obtained due
to the frequency dependence of the self energy
$\Sigma_{\alpha\alpha}(\omega)$. A broad band adds to the
unbroadened  zero phonon line. Its weight with respect to the
total absorption area is $Z_\alpha=1/(1+S_\alpha)$, with the
Huang-Rhys factor $S_\alpha$ stemming from the self energy
derivative
\be\label{Huang-Rhys} S_\alpha = -\left.\frac{d {\mathrm Re}
\Sigma_{\alpha \alpha}(\omega)}{d\omega}\right|_{\omega=\ea-i0} =
\inte dE \, n(E)\, \frac{J_{\alpha\alpha}^\alpha(E)}{E^2} > 0 \,.
\ee
This result for $Z_\alpha$ agrees up to first order in $S_\alpha$
with the exact result from the independent Boson model, $Z_\alpha
= \mathrm{exp}(-S_\alpha)$. However, the single-state solution
neglects completely the inter-state scattering which leads to the
Lorentzian broadening of the lines.

\subsection{Semi-Markov approximation}\label{sec:theory_semi}

In the full non-Markovian solution, both the Lorentzian
broadening of the ZPL and the  broad bands are obtained using the
full self energy \Eq{self_def}. However, the numerical  cost of
$O(N^3)$ coupling functions makes this treatment unfeasible for the
large number $N$ of excitons states covered by the focus of a typical
near-field experiment. Thus, either one has to select  fewer
states, or one has to resort to another approximation which still
preserves the main features of the absorption spectrum.

Aiming at this goal, we introduce the ``Semi-Markov
approximation" which neglects the off diagonal terms in the self
energy and keeps the frequency dependence only in the fully state
diagonal terms:
\bean \label{SM_self} \Sigma_{\alpha\eta}(\omega) & \rightarrow
& \delta_{\alpha\eta} \Sigma_\alpha^{SM}(\omega) \, ,\\ \nonumber
\Sigma_\alpha^{SM}(\omega) \! & \equiv & \!
  \int\!dE\,  \frac{n(E)\,J_{\alpha\alpha}^\alpha(E)}{\omega -\ea +E} \\ \nonumber
&+& \sum_{\beta\neq\alpha} \int\! dE\,
\frac{n(E)\,J_{\alpha\alpha}^\beta(E)}{\ea-i0-\eb+E} \,.\\
\eea
A linear expansion of $\Sigma_\alpha^{SM}(\omega)$ around the
exciton pole $\ea$  shows (cf. App.\,B) that the ZPL has a
reduced width
\be \label{gamma_red}
\gamma_\alpha^{SM} = \gamma^M_\alpha/(1+S_\alpha) \,.
\ee
The semi-Markov approximation is acceptable as long as the off
diagonal elements of the self energy are small with respect to
the differences between diagonal elements of the matrix problem
\Eq{lin_sys}. Since these are dominated by the exciton
eigenergies, the semi-Markov approximation is expected to deviate
markedly from the full non-Markovian treatment only in spectral
regions with a high density of overlapping states. Although much
less demanding in computer power (only $(N^2+N)/2$ coupling
functions needed - as in the Markov approximation), both broad
bands and broadening of the ZPLs are obtained in the semi-Markov
approximation.
\begin{table}[b]
\caption{\label{tab_mat_const}Material constants for bulk GaAs
and other parameters used in the numerical simulations. }
\begin{ruledtabular}
\begin{tabular}{llcc}
longitudinal sound velocity & $u_S$      & 5.33  & nm/ps  \footnotemark[1]      \\
bulk mass density           & $ \rho_M$ & 5.37 & g/cm$^3$ \footnotemark[1]\\
def. pot. electrons         & $D_c$    & -7000 & meV  \footnotemark[1]    \\
def. pot. holes             & $D_v$    & +3500 & meV  \footnotemark[1]    \\
exciton mass           & $ M $ &   0.300   & $m_0$  \footnotemark[2]\\
in-plane electron mass      & $ m_e$ &   0.067 & $m_0$\\
in-plane hole mass      & $m_h$ &   0.233 & $m_0$  \footnotemark[2]\\
QW width                    & $L_z$     &   5   & nm \\
disorder strength           & $\sigma$ & 4.75  & meV \\
correlation length          & $a_B$ & 9.9 & nm \\
grid size                   & $\Delta_x$& 1.65 & nm \\
\end{tabular}
\end{ruledtabular}
\footnotetext[1]{After Ref.\cite{sian01}}
\footnotetext[2]{After Ref.\cite{siar00}}
\end{table}
\bfi[t]
\centering
\includegraphics*[width=8.8cm]{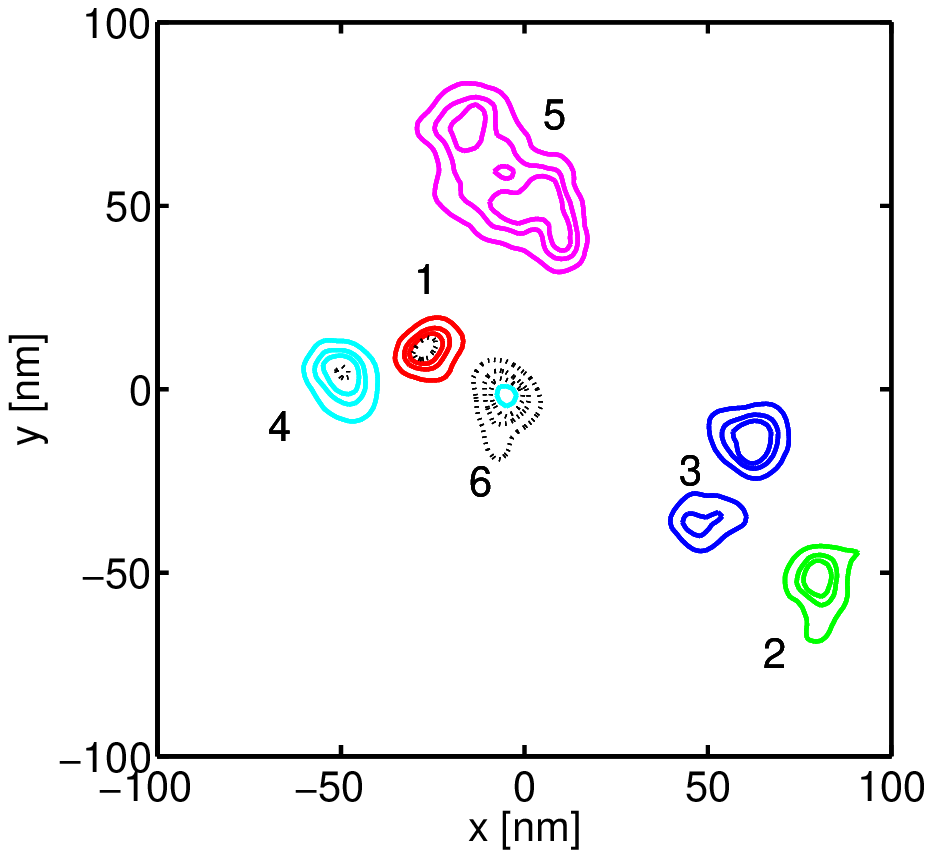}
\caption{\label{im:fig2}(Color online) Probability amplitude of six exciton COM
wave functions in a 5\,nm wide GaAs/AlGaAs QW. The lines are
drawn at 50, 30, and 10\%  of the peak probability of each wave
function. Notice that state 1 and state 6 do overlap.} \efi

\section{\label{sec:results} Results}

We present now numerical solutions of the theory developed in
\Sect{sec:theory}. The Schr\"odinger equation \Eq{COMmotion} has
been discretized on a square grid with step size $\Delta_x$. An
uncorrelated potential corresponding to $(dE_a/dL_z)\Delta
L_z(\R)$ in \Eq{V_def}  has been generated and convoluted with
the relative exciton wave function, using parameters of a 5 nm
wide GaAs/Al$_{0.3}$Ga$_{0.7}$As QW. The resulting disorder potential $V(\R)$ is
computed on the grid and has a correlation length close to the
exciton Bohr radius $a_B$ of this specific QW. The variance
$\sigma$ of this correlated potential has been adjusted to fit
experimental results from the speckle analysis \cite{koch02a}.
The diagonalization of \Eq{COMmotion} on a simulation mesh of
$128\times128$ grid points with periodic boundary conditions was
achieved via an ARPACK-based package. All used material constants
and simulation parameters are listed in \Tab{tab_mat_const}.  In
particular the in-plane hole mass is taken from an excitonic
$k\cdot p$ calculation by Siarkos \etal \cite{siar00} which takes
into account the heavy-light hole mixing. An effective hole mass
of $m_h = 0.233\,m_0$ could be extracted which describes
satisfactorily the exciton COM motion for a wide range of quantum
well thicknesses. We have used this hole mass for the
effective-mass calculation in the present paper.

In \Fig{im:fig2} we display the amplitudes $|\psia|^2$ of six
computed states, whose eigenergies are contiguous. These states
were selected by purpose of exhibiting various combinations of
oscillator strength and spatial overlap to other states. In
particular both quantities are large for $\psi_1(\R)$ and both
are small for $\psi_3(\R)$. State $\psi_5(\R)$ has large
oscillator strength and small overlap, while for $\psi_6(\R)$ the
contrary holds. Here, we are restricted to six states only since
we are going to compare the computationally most demanding
spectra using the full self energy matrix \Eq{self_def} with
other approximations.
\bfi[t]
\centering
\includegraphics*[width=8.8cm]{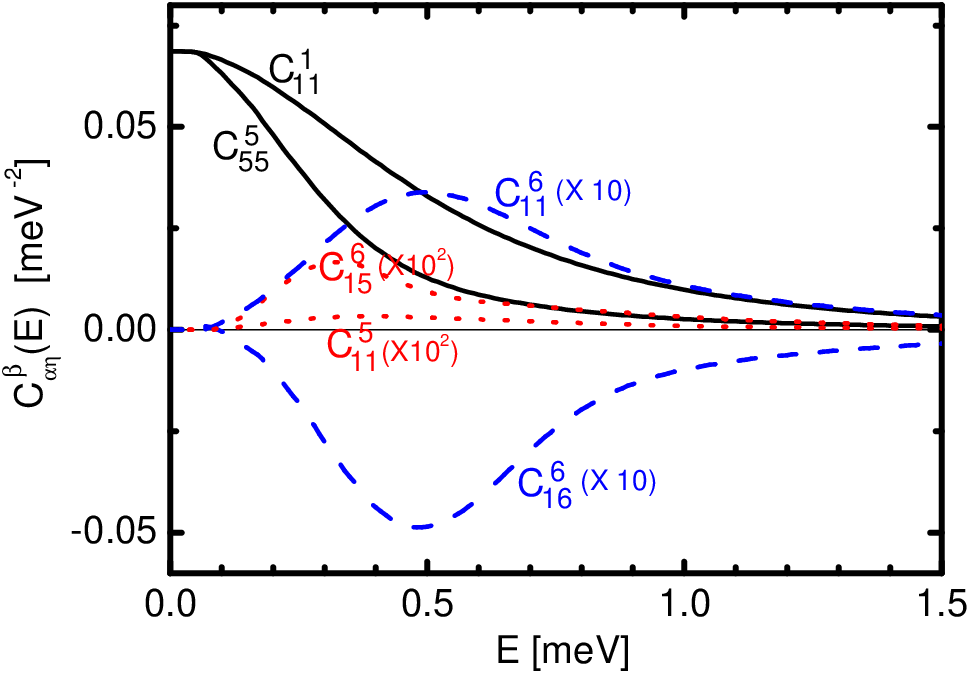}
\caption{\label{im:fig3}(Color online) Selected components of the
coupling function $C_{\alpha\eta}^\beta(E) \equiv
J_{\alpha\eta}^\beta(E)/E^3$ for some of the states of
\Fig{im:fig2}. The off-diagonal components have been magnified by
the factors given in brackets.} \efi

The coupling function $J_{\alpha\eta}^\beta(E)$ is formed using
the eigenstates $\psia$ and represents the most time-consuming
part of the numerical task.  The fully diagonal components
$J_{\alpha\alpha}^\alpha(E)$ are responsible for pure dephasing
in the single-state limit. These components, divided  by $E^3$,
are approximately Gaussians with FWHM inversely proportional to
the localization length  $\Lambda_\alpha$ of state $\alpha$ (full
curves in \Fig{im:fig3} for $\alpha=1$ and $\alpha=5$). Their
localization lengths (calculated from the inverse participation
ratio) are $16.4\,$ and $36.9\,$nm, respectively. Partially
off-diagonal terms $J_{\alpha\alpha}^\beta(E)$ with
$\beta\neq\alpha$ appear in the Markovian scattering rates
$\gamma_\alpha^M$ (\Eq{markov_gamma}) and are sensitive to the
spatial overlap between $\psia$ and $\psib$. In this sense,
$\psi_1(\R)$ overlaps with $\psi_6(\R)$ about 100 times more
effectively than with $\psi_5(\R)$, according to \Fig{im:fig3}.
Finally, fully off-diagonal terms $J_{\alpha\eta}^\beta(E)$ with
$\alpha$, $\beta$, $\eta$ all different account for deviations
between the full solution with self energy \Eq{self_def} and the
semi-Markov level with self energy \Eq{SM_self}.  All nondiagonal
terms of the coupling function vanish at $E \rightarrow 0$  with
power $E^4$ due to the orthogonality of the exciton wave
functions.
\bfi[t]
\centering
\includegraphics*[width=8.8cm]{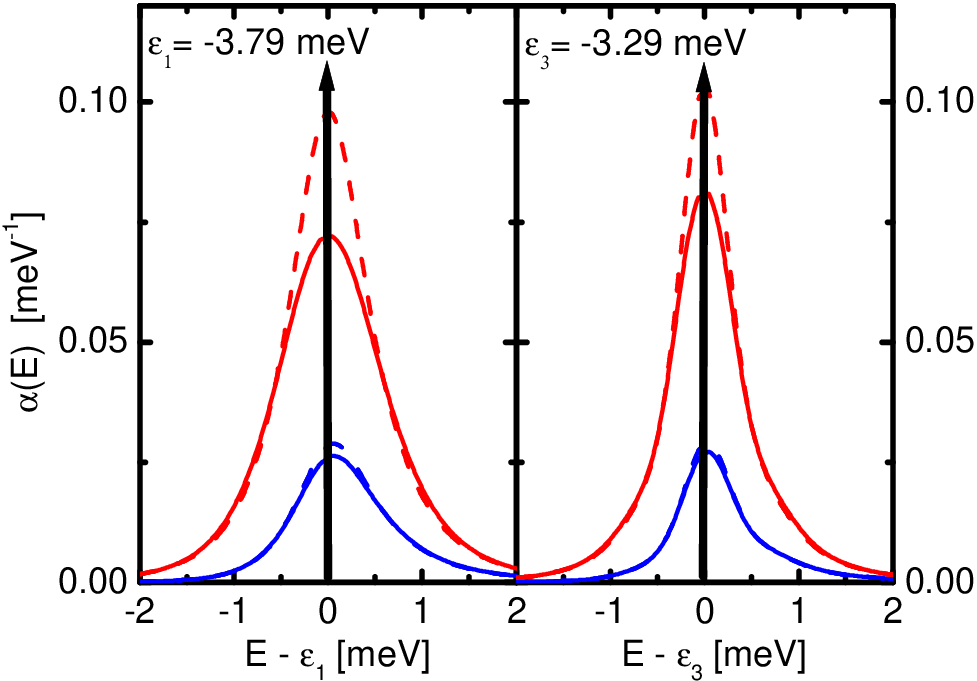}
\caption{\label{im:fig4}(Color online) Absorption lineshape for a
single state at $T=40\,$K (upper curves, red) and  $T=10\,$K
(lower curves, blue). Results from the 2nd Born approximation
(solid lines) are compared with the exact solution (dashed) for
state $\alpha=1$ (left panel) and $\alpha=3$ (right panel) of
\Fig{im:fig2}.  The localization lengths are $\Lambda_1=16.4\,$nm
and $\Lambda_3=23.3\,$nm. The arrows indicate the ZPLs, which
correspond also to the Markov limit of the absorption spectra.}
\efi

Having computed the coupling functions, the self energy at the
desired level of approximation (single state, Markov,
semi-Markov, or full non-Markov ) can be easily generated. Using
this input, the Green's function and the absorption spectrum are
obtained. First we have checked the quality of our 2B treatment
against the exact solution (Independent Boson Model) which is
available in the single-state case only. In \Fig{im:fig4} we show
the absorption spectra for state $\alpha=1$ and $\alpha=3$ of
\Fig{im:fig2}. The broad band stems from phonon absorption (left
side of the ZPL) and phonon emission (right side). Therefore, the
asymmetry evolves into a more symmetric shape as temperature is
raised. There is very good agreement between 2B and exact results
at $T=10\,$K. However, at higher temperatures the broad band is
underestimated in the self energy approach.

In \Fig{im:fig5} we display the absorption spectrum for the system
of six exciton sates. The Markovian spectrum (dashed) is
simply a superposition of Lorentzian lines, whose widths are set
by  Fermi's golden rule, \Eq{markov_gamma}. Since the deformation
potential matrix elements $\tabq$ are sensitive to the spatial
overlap of the COM wave functions (see \App{sec:app_me}), states
2,3,5 exhibit narrower lines compared to states 1,4,6 which have
a stronger overlap.
\bfi[t]
\centering
\includegraphics*[width=8.7cm]{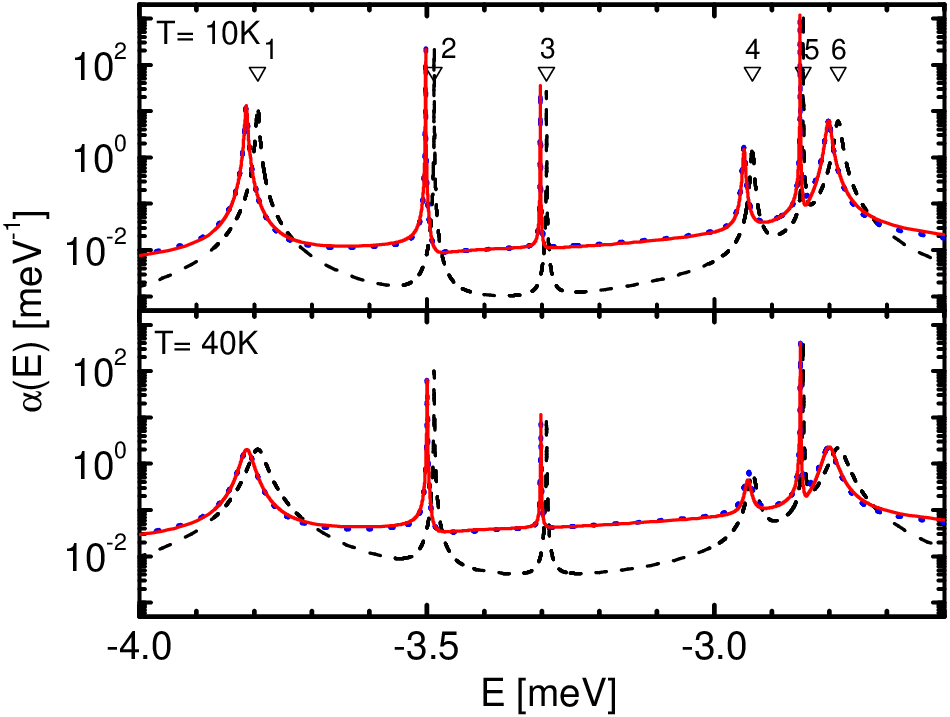}
\caption{\label{im:fig5}(Color online) Absorption spectrum
calculated in the Markov approximation (black dashes), with
semi-Markov quality (blue dots), and with full non-Markovian
quality (red full line) at lattice temperatures of $T=10\,$K (top
panel) and $T=40\,$K (bottom panel). Notice the good agreement of
semi-Markov and full non-Markovian results. No focus function is
included. The triangles indicate the positions of the bare
exciton eigenergies.} 
\efi
The full non-Markovian spectrum is characterized by enhanced
absorption between the single lines, as stressed by the
logarithmic representation of  \Fig{im:fig5}. Furthermore a shift
of the peak energies is observed (polaron shift). In the low
energy part of the spectrum ($E\lesssim -3$\,meV) the energetic
spacing among the states is comparable to or larger than the
width of diagonal elements of the coupling function (cf.
\Fig{im:fig3}). Thus, sidebands are clearly noticed although they
overlap.  They are about one order of magnitude larger than the
superposition of the Lorentzian tails in the Markovian spectrum.
Conversely, the energetic distance between states 5 and 6 is much
smaller than the width of the corresponding coupling function
$J_{\alpha\alpha}^\alpha(E)$. In this region the inter-peak
absorption is dominated by the Lorentzian tails of the ZPLs, and
the individual broad bands cannot be easily resolved. A careful
inspection reveals that the Lorentzian peak widths are smaller
than in the Markovian case. According to \Eq{gamma_red} and
\Eq{Huang-Rhys}, the width reduction scales with the Huang-Rhys
factor and is more important for stronger localized states and
higher temperatures, as observed (cf. e.g. the peak of state 1).
In \Fig{im:fig5} we also check the quality of the semi-Markov
approximation against the full solution. The agreement is
excellent at both $T=10\,$K and $T=40\,$K. Thus,  pure dephasing
is dominated by state-diagonal processes. A similar conclusion
has been obtained in Ref.\,\onlinecite{tak99}.
\bfi[t]
\centering
\includegraphics*[width=8.8cm]{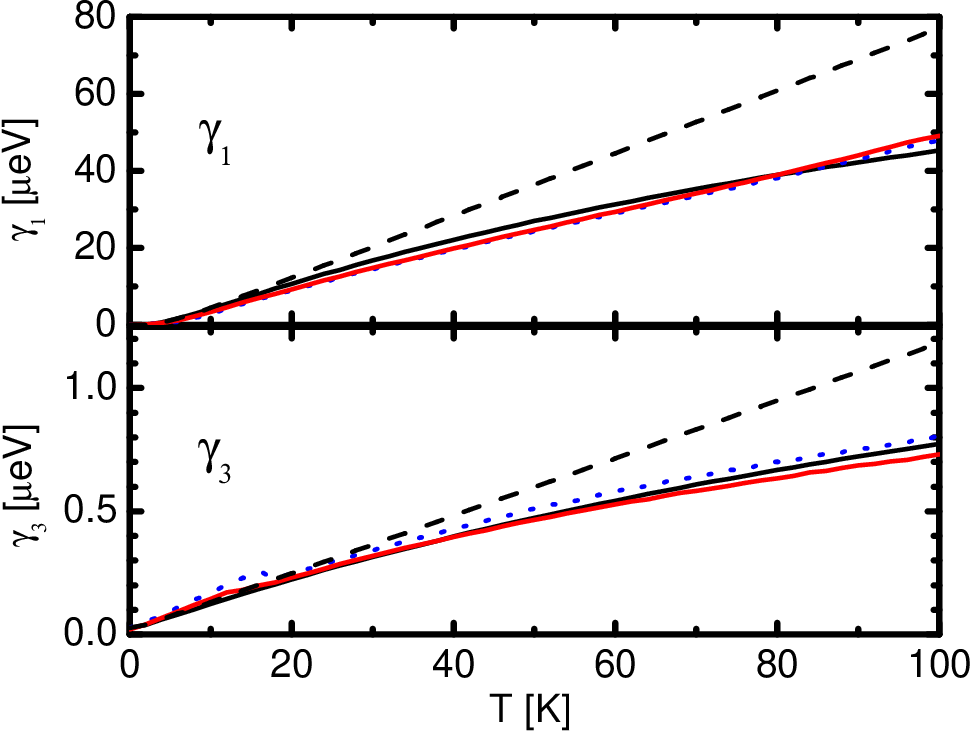}
\caption{\label{im:fig6}(Color online) Temperature dependent ZPL
widths $\gamma_\alpha$ for  lines $1$ and $3$ of the full
non-Markovian (solid red lines) and the semi-Markov spectrum
(dotted blue lines) of \Fig{im:fig4}.  The simplified expression
\Eq{gamma_red} (black solid lines) and the Markov rates
$\gamma_\alpha^M$ (black dashed lines) are also displayed  for
comparison.} \efi

In \Fig{im:fig6} the temperature dependence of the linewidths of
two selected states of \Fig{im:fig5} is studied. The reduced
Markov width \Eq{gamma_red} fairly describes both full solution
and  semi-Markovian widths, as extracted from Lorentz fits in a
narrow spectral window around the ZPLs. This linewidth reduction
from  Markov to non-Markov is a general feature and appears as
soon as the spectral weight of the quasi particle is reduced.
However, being a higher order effect in the coupling strength,
one may ask how an inclusion of higher order terms would alter
this finding. A consequent treatment of all diagrams up to 4th
order in the cumulant expansion by Muljarov \etal
\cite{muljarov05} has shown that the reduction is indeed a
pertinent feature. Its numerical value, however, might be
quantitatively not correct when restricting to the level of 2nd
Born approximation. In  \Fig{im:fig6}, a sublinear dependence of
the linewidth on temperature is found for both states. This can
be easily explained by noting that the Huang-Rhys factor
increases nearly linear with $T$ (cf. \App{sec:app_zpl}).

\bfi[t]
\centering
\includegraphics*[width=8.8cm]{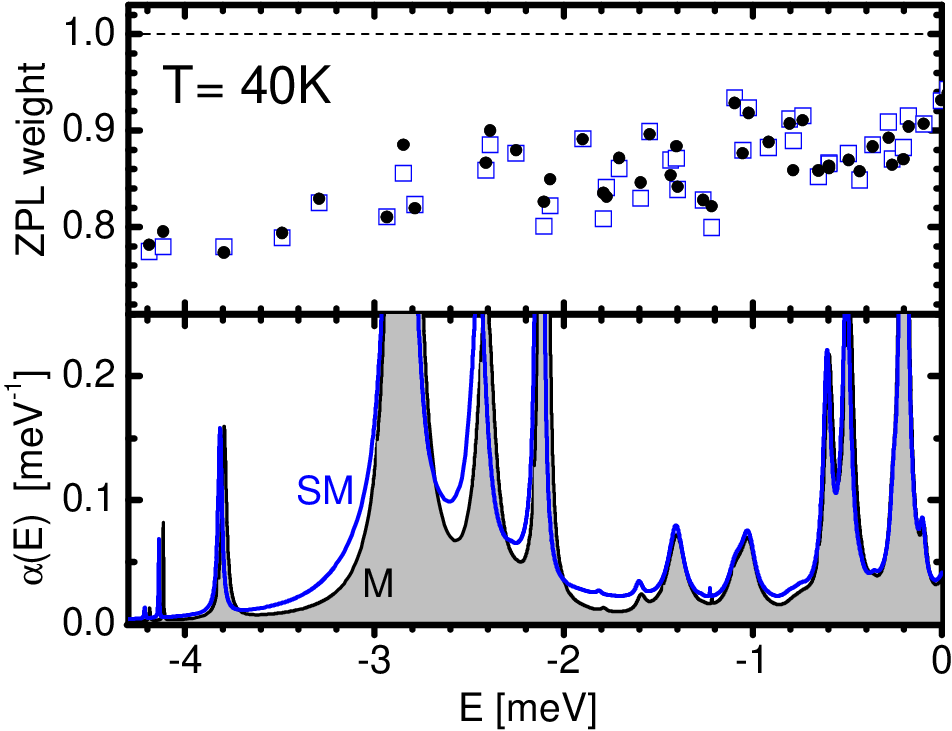}
\caption{\label{im:fig7}(Color online) Top panel: ZPL weights
$Z_\alpha=1/(1+S_\alpha)$ (blue squares) and
$\Lambda_\alpha/(\Lambda_\alpha + bT)$ ratio (black circles) for
a set of 50 exciton states at $T=40\,$K. A good agreement is predicted from \Eq{Za_approx}
 and it is found using  $b=0.12\,$nm/K.
 The dashed line is the constant ZPL weight within the Markov approximation. Bottom
Panel: Markovian (shadowed area) and semi-Markovian (blue solid
line) absorption spectrum for the same set of states at $T=40\,$K
using a focus of $50\,$nm.} \efi

Having tested the features of the semi-Markov approximation in
\Fig{im:fig5} and \Fig{im:fig6}, we apply it to a system of many
exciton states. To come closer to the experimental situation of a
near-field measurement, we have included a Gaussian focus of
50\,nm  width (FWHM of  field intensity). Using the 50 lowest
energy eigenstates of a disorder realization for a 5\,nm wide QW
we obtain the spectrum in \Fig{im:fig7} up to $E=0\,$meV which is
the exciton energy in an ideal QW of the same width. In a near
field geometry, the visibility of a state depends not only on its
oscillator strength, but on the overlap with the illumination
spot as well, as seen from \Eq{ma_fact}. Thus, several states in
the range $-3\,$meV$<E<-2\,$meV appear especially strong in the
present simulation.  The absorption background between the peaks
can be clearly seen on a linear scale in this spectral region. In
the high energy tail ($E\approx 0\,$meV) there is practically no
difference between Markov and semi-Markov spectra. Even on a
logarithmic scale both spectra are nearly identical (not shown).
These observations can be explained considering the localization
length $\Lambda_\alpha$ and the spectral separation of the
exciton states.  $\Lambda_\alpha$ is increasing across the
spectrum from $\Lambda_2=17.2\,$nm (the ground state is dark in the actual illumination geometry) 
up to $\Lambda_{50}=54.1\,$nm.
Via the Huang-Rhys factor,  this leads to an increasing ZPL
weight (cf. \App{sec:app_zpl}) which goes from $Z_2=0.77$ up to
$Z_{50}=0.92$, as seen in the top panel of \Fig{im:fig7}. Thus,
the importance of the non-Markovian broad bands (proportional to
$1-Z_\alpha$) decreases with photon energy.  However, the
observability of this effect strongly depends on the spectral
separation of the resonances, as discussed for  \Fig{im:fig5}.
The spectrally close doublet at about $-4.2\,$meV e.g. does not
show a distinct broad band, in spite of having a short
localization length (large $S_\alpha$).  Thus, for observing
non-Markovian effects in the absorption spectrum from QWs, it is
necessary to position the near-field focus in such a way that
states with small localization length and large spectral
separation are excited.

A non-Markovian computation of the absorption which includes all
optically active exciton states is at the moment beyond our
computational possibilities. We expect somewhat broader ZPL peaks
since more states would be available for phonon-mediated
scattering. However, the broad band absorption between peaks is
not expected to be modified as it is due to intra-state
relaxation (first term on the rhs of \Eq{SM_self}). Thus, a
calculation of the absorption spectrum using all states will not
reveal  new features as compared to what we display in
\Fig{im:fig7}, at least within the semi-Markov approximation.

\section{\label{sec:conclusions} Conclusions}

In conclusion, we have developed a theoretical description of the
exciton-acoustic phonon scattering in disordered QWs which goes
beyond the Markov approximation. We have considered deformation
potential coupling with acoustic phonons and realistic exciton
states resulting from a microscopic simulation. A second order
self energy approach is used for computing the absorption
spectrum from the exciton propagator. At the Markov level the
absorption is a superposition of Lorentzian peaks with widths
defined by Fermi's golden rule, which results from
phonon-assisted scattering between different states. At the full
non-Markovian level, virtual scattering events on the same state
are considered as well (pure dephasing). These processes manifest
itself in broad absorption bands around Lorentzian peaks (ZPL)
whose widths are smaller than according to Fermi's golden rule.
We have also discussed a semi-Markov approximation in which a
diagonal self energy is employed, but the important frequency
dependence of the diagonal term is preserved. This allows to reduce
the numerical task from $O(N^3)$ to $O(N^2)$, where $N$ is the
number of exciton states, while recovering all the important
features of the complete  solution. We have shown that the
non-Markovian effect increases with a shorter localization length
of the exciton wave function, and for low spectral density of
states. A near-field setup could demonstrate these predictions
experimentally.

\begin{acknowledgments}
Support from DFG in the frame of Sfb 296 is gratefully
acknowledged. G.~M. is thankful to Egor Muljarov (Berlin) for
discussions and to Roberto Cingolani (Lecce) for support from
CNR.
\end{acknowledgments}

\appendix
\section{\label{sec:app_me} Interaction matrix elements and coupling function}

The matrix element for optical interband transitions between an
exciton wave function $\Psi_\alpha(\r_e \r_h)$ and a light field
with momentum  $\q$ and field envelope $E(\r)$ is given by
\be\label{dip_me_def} m_\alpha = \mu_{\mathrm{cv}} \inte d\r\,
e^{-i\q\cdot\r}\, E(\r)\, \Psi_\alpha (\r,\r) \ee
with the interband dipole moment $\mu_{\mathrm{cv}}$ of the band
edge states. The factorization \Eq{FactorWF} and transmission
normal to the QW plane  ($\qp=0$) results in
\be\label{ma_fact}
 m_\alpha = \mu_{\mathrm{cv}} \phi_{1s}(0)  O_{eh}  \inte d\R\, E(\r)\, \psia \,,
\ee
with $\phi_{1s}(0)$ as the relative exciton wave function at
the origin, and $O_{eh}$ the confinement overlap integral $O_{eh}
= \inte dz\, u_e(z) \,u_h(z)$.

Within the same factorization Ansatz, the deformation potential
matrix element of Ref.\,\onlinecite{tak85} takes the form
\bean \label{t_ab_q} t_{\alpha\beta}^\q  & =&
\sqrt{\frac{\hbar\wq}{2 u_S^2 \rho_M \, V }} \, (\psi_\alpha
\psi_\beta)_\qp \times \\ \nonumber
& & \Big[ D_c K_e (q_z) \chi( \qp/\eta_e) - D_v K_h (q_z) \chi(\qp/\eta_h) \Big] \\
\eea
with deformation potential constants $D_a$,  sound velocity
$u_S$,  mass density  $\rho_B$, and  normalization volume $V$.
Further, we have defined Fourier transforms of the squared
confinement and relative wave functions, and the COM overlap
\bean
    K_a(q_z) &=& \int \! dz\, u_a^2(z)\, e^{-i q_z z} \nonumber \\
  \chi (\q_\parallel) &=& \int \! d\ro\, \phi_{1s}^{2}({\mathbf{\rho}})\,
    e^{-i\q_\parallel \cdot {\mathbf{\rho}}} \\
    (\psi_\alpha \psi_\beta)_\qp &=& \inte d\R\, \psia e^{-i\qp\cdot\R} \psib
    \,.\nonumber
\eea
In particular, for Gauss confinement states and a hydrogen-like
relative motion we obtain:
\bean
 K_{a}(q_z) &=& \mathrm{exp}\left(-\half(q_z L_a)^2\right) \\
    \chi (\q_{\parallel }) &=&  \left( 1 +
(q_\parallel a_{B}/2)^2 \right)^{-3/2} \, ,
\eea
where $L_e=1.69\,$nm and $L_h=1.34$\,nm. These two values result
from a variational determination of the confinement levels in the
5 nm wide GaAs/AlGaAs QW.  Other material constants for GaAs are
listed in \Tab{tab_mat_const}.

Since the exciton states $\psia$ are numerically generated in
Cartesian coordinates, we proceed with fast Fourier
transformation along $q_x$, $q_y$, $q_z$ and integrate
accordingly \Eq{J_def} as
\bea\label{J_cartesian}
\nonumber J_{\alpha\eta}^{\beta}(E) &=& \frac{V}{(2\pi)^3}
\int\! dq_x dq_y dq_z \, t_{\alpha\beta}^\q t_{\eta\beta}^{-\q} \, \times\\
&& \delta(|E|-\hbar u_S \sqrt{q_x^2+q_y^2+q_z^2} \, .) \eea
The delta function is used to integrate over $q_z$. Special care
is taken to deal with the resulting square root singularities at
the remaining integration limits. Each matrix element $t$
contains a square-root of phonon energy, thus
$J_{\alpha\eta}^{\beta}(E)$ starts at least with a power of
$E^3$.

\section{\label{sec:app_zpl}ZPL width and weight}

In the semi-Markov approximation, the absorption is given by
\be\label{SM_absorption} \alpha(\omega) = \mathrm{Im} \sum_\alpha
\frac{|m_\alpha|^2} {\omega - \ea - \Sigma_\alpha^{SM}(\omega)} \,. \ee
For evaluating the ZPL width, we expand in \Eq{SM_absorption} the
self energy around the pole $\ea$. The real part at the resonance
(polaron shift) $\mathrm{Re}\Sigma_\alpha^{SM}(\ea) $ is small
and can be dropped, but its derivative has to be kept as
\be -\left. \frac{d}{d\omega} \mathrm{Re}
\Sigma_\alpha^{SM}(\omega)\right|_{\omega = \ea-i0}  =  S_\alpha
\, . \ee
The imaginary part is taken as $
\mathrm{Im}\Sigma_\alpha^{SM}(\ea-i0)  = \gamma^M_\alpha/2$.
Putting this into \Eq{SM_absorption} we end up with an absorption
lineshape
\be \alpha(\omega\approx\ea) = \sum_\alpha \frac{|m_\alpha|^2}{1+S_\alpha} \,
\mathrm{Im} \frac{1}{\omega -\ea - i \gamma_\alpha^{SM}/2} \,,
\ee
which has a reduced ZPL weight of $1/(1+S_\alpha)$ and a reduced
width $\gamma_\alpha^{SM}= \gamma_\alpha^{M}/(1+S_\alpha)$.

At high temperatures, the phonon occupation can be replaced by
$n(E) \rightarrow E/k_B T$. Together with the Gaussian shape of
the diagonal coupling factor (see \Fig{im:fig3}) we can evaluate
the Huang-Rhys factor \Eq{Huang-Rhys} in closed form
\be S_\alpha \approx \inte dE \frac{k_B T}{E} E^3 A\,
\mathrm{exp}\left[- \left(\frac{\Lambda_\alpha E}{
u_s}\right)^2\right] \frac{1}{E^2} = b \frac{T}{\Lambda_\alpha}
\, . \ee
Thus, the ZPL weight is approximately given by
\be\label{Za_approx}
Z_\alpha = \frac{1}{1+S_\alpha} \approx \frac{\Lambda_\alpha}{\Lambda_\alpha+ bT} \,.
\ee
This approximation is getting better for less localized states, which
can be seen in the top panel of \Fig{im:fig7}.

\bibliography{PRB}

\end{document}